# Performance of Macro-Scale Molecular Communications with Sensor Cleanse Time

Siyi Wang, Weisi Guo[†], Song Qiu[†], Mark D. McDonnell
Institute for Telecommunications Research, University of South Australia, Australia
[†]School of Engineering, The University of Warwick, United Kingdom
siyi.wang@mymail.unisa.edu.au, [†]{weisi.guo, song.qiu}@warwick.ac.uk, mark.mcdonnell@unisa.edu.au

***Abstract*—In this paper, we consider a molecular diffusion based communications link that conveys information on the macro-scale (several metres). The motivation is to apply molecular-based communications to challenging electromagnetic environments. We first derive a novel capture probability expression of a finite sized receiver. The paper then introduces the concept of time-aggregated molecular noise at the receiver as a function of the rate at which the sensor can self-cleanse. The resulting inter-symbol-interference is expressed as a function of the sensor cleanse time, and the performance metrics of bit error rate, throughput and round-trip-time are derived. The results show that the performance is very sensitive to the sensor cleanse time and the drift velocity. The paper concludes with recommendations on the design of a real communication link based on these findings and applies the concepts to a test-bed.**

## I. Introduction

There exists a need to convey information over *free space* in both human society and in animal populations. This occurs on a body-to-body level (macro-scale) and on a cell-to-cell level (micro-scale). Over the millenniums, more complex systems have been built to transmit continuous data streams, as opposed to discrete signals.

*1) In Nature: Chemical Signalling:* Molecular communications is a nature-inspired concept that builds on existing biological systems. Molecular signalling does not just occur on a microscopic scale in between bacteria, but also over longer distances outdoors between members of the species. For example, pheromones are used for long range signalling between social insects, fish, and elephants [1]. Usually the data transported on these channels are a small finite set of distinctive messages that are encoded by chemical compositions.

*2) In Engineering: Molecular Communications:* The proposed system is different to nature in the respect that we transport a *reliable continuous stream of data* using molecules, which is far more challenging [2]. The motivation behind using molecules to carry information lies in challenging electromagnetic (EM) propagation environments. In many industrial applications, it is desirable to convey information without tethers. Sensors are typically in embedded locations and are attempting to transmit information across complex environments. The challenges that face EM wave based systems include: i) antenna size need to be proportional to wavelength; ii) EM-waves suffer high absorption losses; iii) EM-waves cannot easily propagate through cavities that do not act as a wave-guide; and iv) transmission is limited by available spectrum and energy. Of course, over the decades of EM-based system research, solutions have been developed to somewhat overcome the above challenges. The authors themselves have prototyped cooperative transmission techniques for such environments [3]. However, one of the main barriers to ubiquitous wireless sensor deployment remains data extraction [4].

For such sensor applications, there is often a requirement to design small sized sensors that can deliver data at very low energy levels. Examples include monitoring corrosion in structures (e.g., bridge casings, pipe and tunnel networks that cannot act as waveguide [5]), and also in areas where one wants to minimise electromagnetic radiation or suffer from excess radiation interference. Molecular communications has been proposed as a solution over the past few years [2], [6], [7], and investigated for different environments in [8]. More recently, the world's first test-bed that can transmit continuous messages has been demonstrated [1]. The open challenges in this area include finding accurate characteristic functions for the impulse response of the channel and understanding the resultant communication performance possibilities.

In the first part of the paper, we derive from first principles a novel captured molecule concentration equation. We compare this to those in existing literature and demonstrate why it might be more suited to molecular communications. In the second part, we derive the resulting bit error probability and throughput rate with due consideration to both inter-symbol-interference (ISI) and Gaussian noise. This is achieved by applying the minimum error probability (MEP) criterion of a standard Bayesian detection framework.

## II. Captured Molecule Concentration

### A. Conditions for Impulse Response Capture

This paper considers molecular communications in pipe networks and requires specific boundary conditions to diffusion equations that are not commonly used in literature. More specifically, we need to incorporate an impulse emission (input), a perfect molecular capturing receiver, and a semi-infinite environment. The Fokker-Planck equation [10] predicts the flux of molecular diffusion as a function of the diffusivity

[1]Kinboshi system developed by Dr. Farsad (York) and Dr. Eckford (York) with Dr. Guo (Warwick) [9]

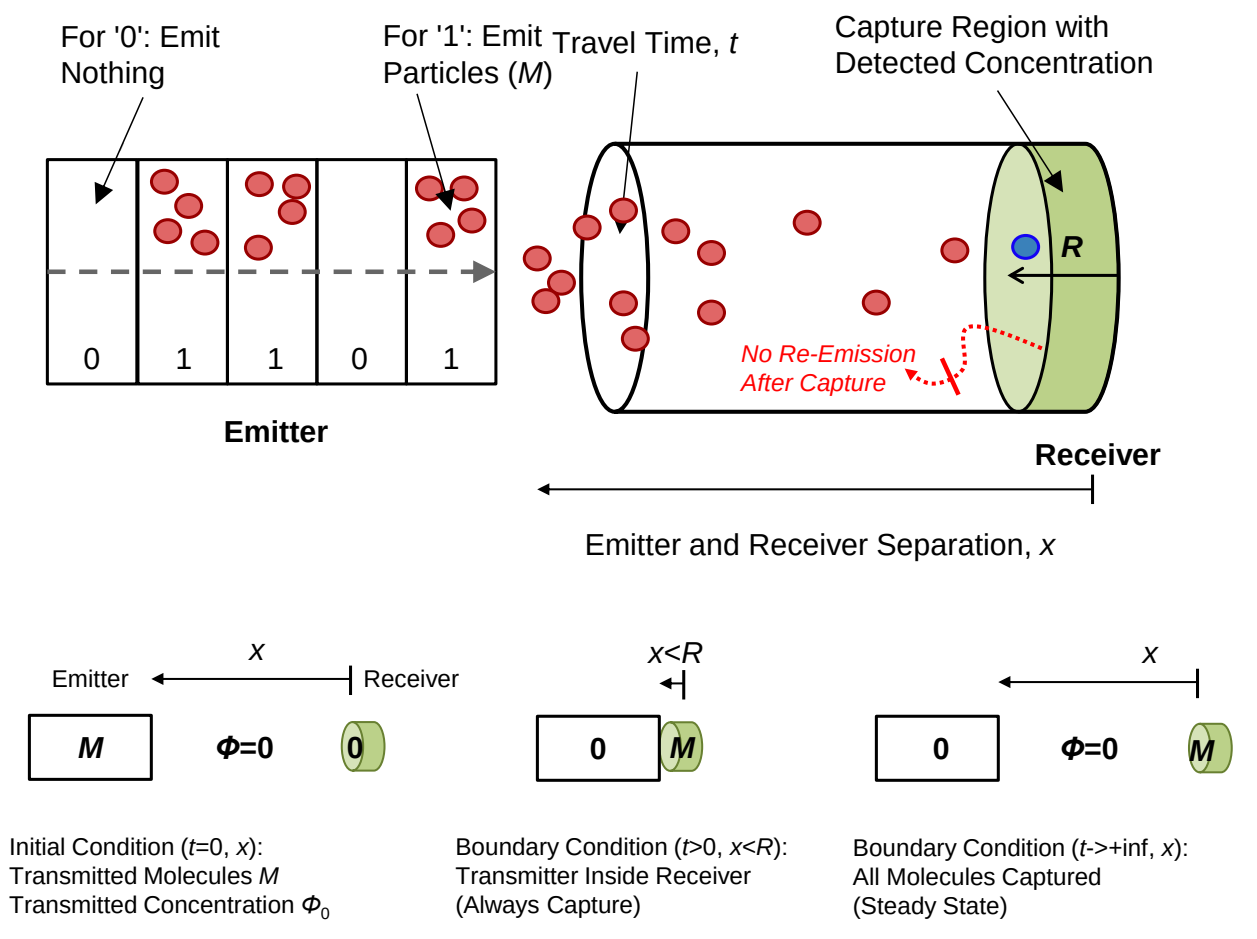

Fig. 1. Illustration of molecular communication link using On-Off-Keying (OOK) modulation scheme with a receiver that has a capture radius of $R$. Three conditions are presented: i) initial pulse transmission ($t = 0$), ii) always full capture ($x < R$), and iii) infinite capture ($t \to +\infty$).

coefficient $D$ ($m^2$/s):

$$\frac{\partial \phi(x,t)}{\partial t} = -v\frac{\partial \phi(x,t)}{\partial x} + D\frac{\partial^2 \phi(x,t)}{\partial x^2}, \tag{1}$$

where $\phi(x,t)$ is the concentration (molecules/m), $t$ is the time elapsed (s), $v$ is the drift velocity (m/s) and $x$ is the distance from source (m). The diffusivity $D$ is based on experimental values [11].

In order to solve the ordinary partial differential equation (OPDE), one initial ($t = 0$) and two boundary conditions must be set. Let us consider the system shown in Fig. 1. The molecular communication link contains a receiver and a transmitter separated by a distance $x$ and the receiver is of size $R$. The concentration $\phi$ is defined as the molecule concentration at any point outside the capture zone (can include the transmitter). The system has several properties: i) the propagation environment is non-infinite, ii) the information transmitted is modulated by concentration (amplitude), and iii) once the receiver detects a molecule, it is fully absorbed and cannot be recycled. Therefore, 3 key conditions are presented that are *necessary* to fully describe the communication system:

1) *Initial Impulse,* $\phi(x,0) = \phi_0$: at $t = 0$, a pulse of concentration $\phi_0$ is emitted at the transmitter which is located at $x$ distance away from receiver;
2) *Capture Zone and No Re-emission,* $\phi(R,t) = \phi_0$: if the transmitter is always located at $x \leqslant R$ from the receiver, anything emitted will be captured immediately. At any time $t > 0$, the concentration outside the zone is $\phi = 0$.
3) *Long-Term Capture,* $\phi(x,\infty) = 0$: if a molecule is captured, it cannot be re-emitted. Therefore, over a long time ($t \to +\infty$), the receiver captures all the molecules and the external concentration is $\phi = 0$.

There are a number of alternative conditions used in other papers, which we will now examine and show how they are *not suitable* for the molecular communication system outlined here:

- *Infinite Source* [12], [13]: this condition states that an infinite source of molecules provide a continuous and finite flux of molecules, such that $\phi(r,t) = \phi_0$.
- *Infinite Environment* [14], [15]: this condition states that the propagation environment is infinite. This is valid for an open atmosphere communication system, but is not realistic for enclosed structural environments that we consider.
- *Fast Sensor Response*: that is to say the molecules at sensors are immediately converted to electrical charge and there is zero aggregated chemical interference from previous emissions. In reality gas phase sensors typically have a response time of several seconds and therefore the response can not be arbitrarily fast.

### B. Captured Concentration

The non-capture concentration ($\phi(x,t)$) equation that satisfies both Fokker-Planck Eq. (1) and the three conditions is:

$$\phi(x,t) = \frac{\mathsf{M}}{\sqrt{\pi D t}} \exp\left[-\frac{(x-R-vt)^2}{4Dt}\right] \quad \text{for } x > R, \tag{2}$$

and $\phi(x,t) = 0$ for $x \leqslant R$. Assuming conservation of molecules, the captured number of molecules is simply $\theta_c = \mathsf{M} - \int_R^x \phi(u,t)\,\mathrm{d}u$. Given that all molecules will be eventually captured (condition 3), the maximum cumulative captured number of molecules is $\theta_c = \mathsf{M}$. Therefore, the *cumulative captured number of molecules* can be defined as the difference between the total emitted number of molecules (M) and the number of molecules outside the capture zone up to any given time $t$. The resulting number of molecules captured is a monotonically increasing function:

$$\begin{aligned} \theta_c(x,t) &= \mathsf{M} - \int_R^x \frac{\mathsf{M}}{\sqrt{\pi D t}} \exp\left[-\frac{(u-R-vt)^2}{4Dt}\right] \mathrm{d}u, \\ &= \mathsf{M}\left[\operatorname{erfc}\left(\frac{v}{2}\sqrt{\frac{t}{D}}\right) - \operatorname{erf}\left(\frac{x-R-vt}{2\sqrt{Dt}}\right)\right], \\ &= \mathsf{M}\left[\operatorname{erfc}\left(\frac{x-R}{2\sqrt{Dt}}\right)\right] \quad \text{for: } v = 0. \end{aligned} \tag{3}$$

The zero-drift case is similar to the theoretical results derived in [12], [16]. The concentration inside the capture zone is simply $\phi_c = \theta_c/R$.

### C. Captured Probability Functions

In order for a receiver to sample at the point where there is the greatest flux of molecules, one has to find the flux (captured molecules per second). The number of captured molecules between any two arbitrary points in time is given by $\theta_c(x,t+\Delta t) - \theta_c(x,t)$, which is not a monotonic function. If one assumes that the M molecules are transmitted independently, the probability of capturing one molecule is given by

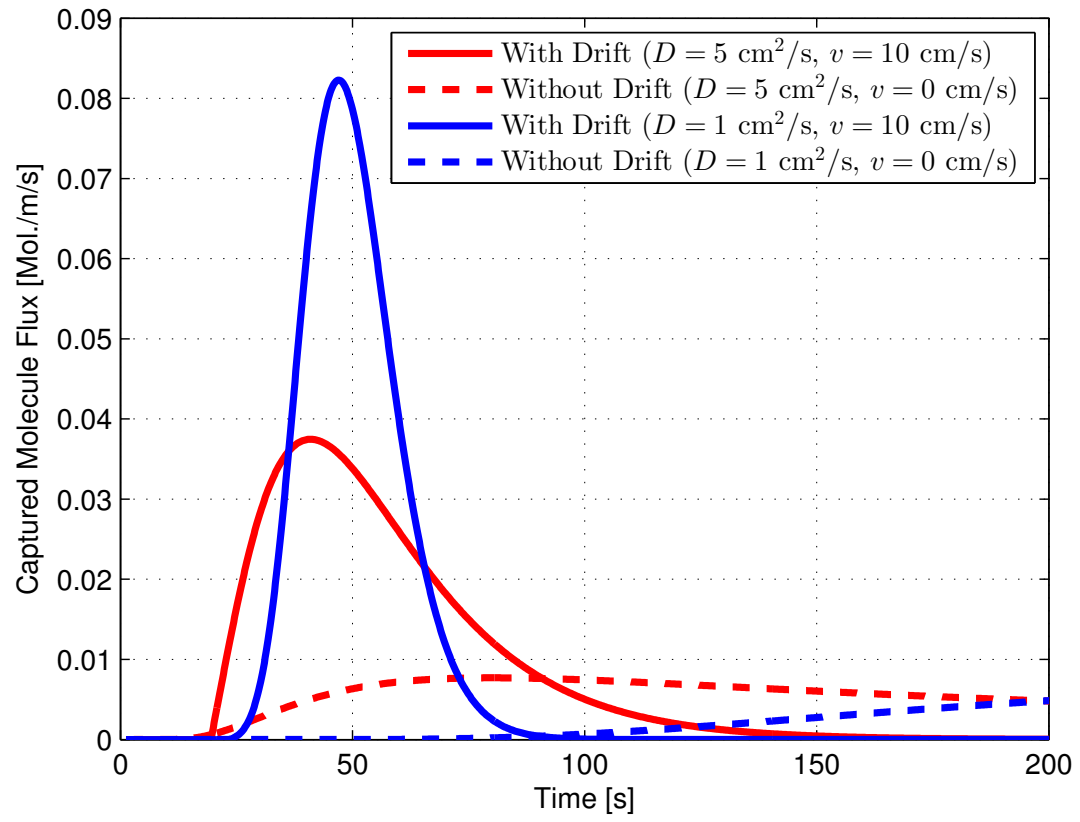

Fig. 2. Captured molecules flux (molecules/m/s) for zero-drift ($v = 0$) and positive drift ($v = 10$ cm/s), for $x = 1$ m and $R = 1$ cm.

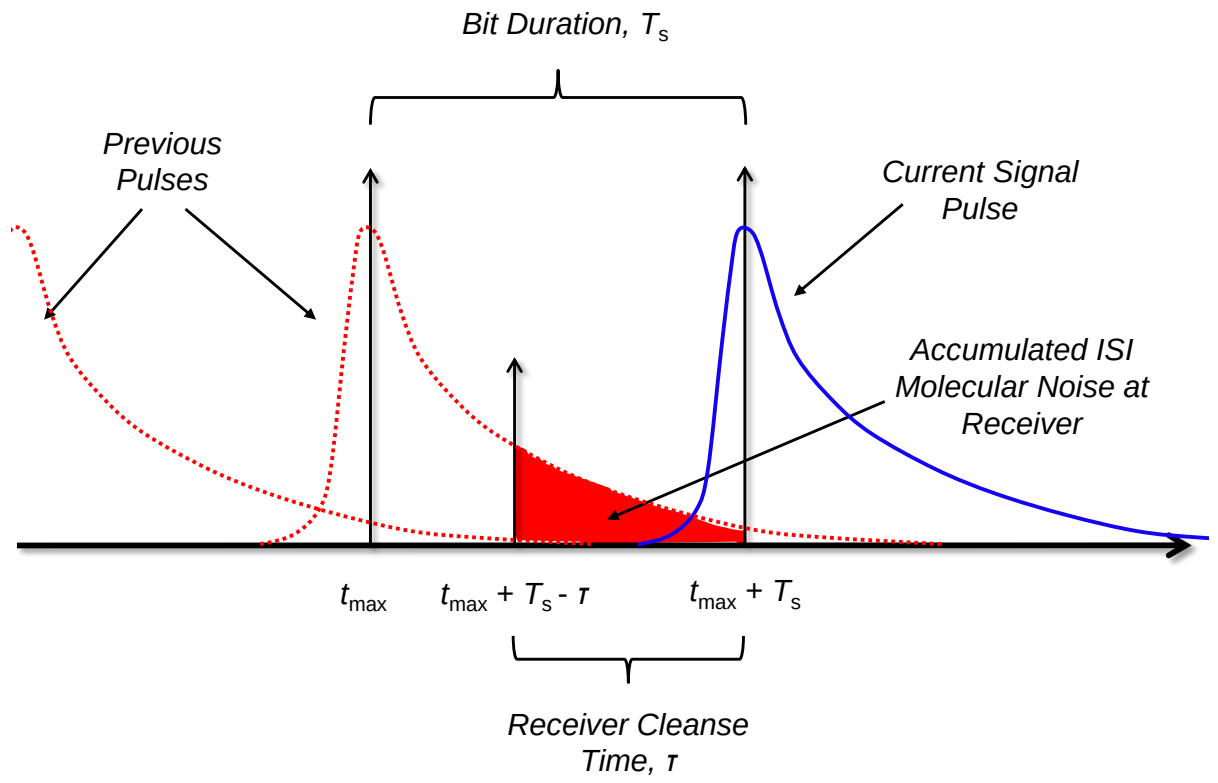

Fig. 3. Illustration of ISI received at the receiver as a result of molecules captured in the previous $\tau$ seconds.

Eq. (3) divided by M. Therefore, the cumulative distribution function (CDF) is: $F_T(t) = \text{erfc}\left(\frac{v}{2}\sqrt{\frac{t}{D}}\right) - \text{erf}\left(\frac{x-R-vt}{2\sqrt{Dt}}\right)$. The partial derivative of the cumulative function with respect to time yields the likelihood of capture between any particular time $t$ and $t + \Delta t$ for a $\Delta t \to 0$:

$$\begin{aligned} f_T(t) &= \frac{\partial F_T(t)}{\partial t}, \\ &= \frac{(x-R+vt)\exp\left[-\frac{(x-R-vt)^2}{4Dt}\right] - vt\exp\left(-\frac{v^2 t}{4D}\right)}{2\sqrt{\pi D t^3}}. \end{aligned} \quad (4)$$

In the special case of no drift velocity, $v = 0$, we have

$$f_T(t) = \frac{(x-R)}{2\sqrt{\pi D t^3}}\exp\left[-\frac{(x-R)^2}{4Dt}\right], \quad \text{for: } v = 0. \quad (5)$$

This can be interpreted as the flux of captured molecules when scaled by $\text{M}/R$. A plot of the captured molecules flux (molecules/m/s) for zero-drift ($v = 0$) and positive drift ($v = 0.1$ m/s) is presented in Fig. 2. It can be seen that a small drift velocity can significantly shorten the peak arrival time and increase the peak-to-average ratio of the received impulse response.

In *turbulent flow*, increasing the diffusivity $D$ will broaden the width of impulse response pulse, but also shorten the arrival time. Fig. 2 demonstrates this by increasing $D$ from 1 to 5 cm$^2$/s. Therefore, turbulence can be beneficial from the point of view of shortening the communication round-trip-time, but will cause greater inter-symbol-interference for the same transmit bit duration. As far as we are aware, both the captured number of molecules and the probability density function (pdf) are novel results that differentiate from existing expressions for the reasons mentioned in Section **??**. By utilising the flux or pdf, the optimal sampling point and the number of captured molecules can be found.

## III. Performance

### A. Optimal Sampling Point

As shown in Fig. 1, the paper considers the binary digital On-Off-Keying (OOK) modulation scheme without forward-error-correction coding. The receiver and the transmitter are assumed to be synchronised and that the receiver detects a flux of captured molecules over a $\Delta t$ period. The maximum flux is given by solving $\partial f_T(t)/\partial t = 0$ for $t$, yielding:

$$t_{\text{max}} = \frac{(x-R)^2}{6D}, \quad \text{for: } v = 0. \quad (6)$$

For a positive drift velocity, only a numerical solution can be found. Note, this is a similar result as that arrived with the non-capture diffusion equations in [15].

Let us assume that the receiver samples over $\Delta t = \tau$ period, where $\tau$ is sufficiently small compared to the diffusion process. By substituting the optimal sampling time into the number of molecules captured Eq. (3), the peak captured concentration is:

$$\begin{aligned} \phi_{\text{max.}} &= \frac{\theta_c(x, t_{\text{max}}) - \theta_c(x, t_{\text{max}} - \tau)}{R}, \\ &= \frac{M}{R}\left[\frac{3D\tau}{(x-R)^2\sqrt{\pi/6}}\right]\exp\left(-\frac{3}{2}\right), \quad \text{for: } v = 0. \end{aligned} \quad (7)$$

For the zero-drift case, it can be seen that the received signal power (captured molecules flux) is a linear function of the diffusivity $D$ and approximately an inverse square relationship with the transmission range $x$. The expected delay (or round-trip-time) of such a system is given by $t_{\text{RTT}} = \frac{(x-R)^2}{3D}$ for a reciprocal channel with zero-drift, which can be interpreted as $1/(3D)$ per metre of transmission distance squared.

### B. Inter-Symbol-Interference (ISI)

Unlike traditional communications with electromagnetic waves, molecular communications cannot effectively alter the shape of the arrival pulse, such that ISI is avoided. Therefore, the molecules from previously transmitted symbols potentially becomes a dominating source of error. Examining more closely, most molecular receivers aggregate molecules for a period $\tau$, after which the molecules dissipate (cleansed). Therefore, the ISI is not only comprised of the aggregated molecules from previous symbols at the sample time $t_{\max}$, but the aggregated molecules from $t_{\max} - \tau$ to $t_{\max}$.

Referring to Fig. 3, let us consider an observed signal pulse which is being optimally sampled at its peak concentration point. It receives ISI from $N \to +\infty$ previous symbols. Let $T_s$ denote the transmit bit duration. Using the time reference of the previous symbol, the aggregated interference is from time $t_{\max} + T_s - \tau$ to $t_{\max} + T_s$. Therefore, the lower-bound to the number of molecules accumulated at the receiver when it is sampling a pulse that is receiving ISI from previous symbols is:

$$\theta_{\text{ISI}} = \sum_{n=1}^{+\infty} \chi_n \big[\theta_c(x, t_{\max} + nT_s) - \theta_c(x, t_{\max} + nT_s - \tau)\big], \tag{8}$$

where $\chi_n$ represents the emitter transmitting a 1 or 0. The total ISI molecules captured by the receiver over time $\tau$ converges absolutely, and more importantly it was found through empirical simulations that it is not dependent on $D$.

The concentration $I$ in the capture zone is defined by $\theta_{\text{ISI}}/R$. The maximum value of ISI ($\theta_{\text{ISI,max}}$) occurs when $\chi_n = 1$. Given equal probability of transmitting a 1 or a 0, the mean is $\mu_{\text{ISI}} = \frac{\theta_{\text{ISI,max}}}{2R}$. The precise variance of the ISI is challenging to find explicitly, so the paper considers the upper-bound variance, which is:

$$\sigma^2_{\text{ISI}} = \mathbb{E}[\theta^2_{\text{ISI}}] - \mu^2_{\text{ISI}} = \frac{\theta^2_{\text{ISI,max}}}{2R^2} - \left(\frac{\theta_{\text{ISI,max}}}{2R}\right)^2 = \mu^2_{\text{ISI}}. \tag{9}$$

### C. Bayesian Decision Threshold

We consider two forms of noise in the system, one from the previously mentioned ISI ($I$) and the other from additive Gaussian noise ($N$) at the receiver's hardware and from background ambient molecules (chemical contamination or interference). Let $\vartheta$ denote the captured concentration including noise:

$$\vartheta = \chi\phi_{\max.} + I + N. \tag{10}$$

The distribution of ISI is given by the pdf of the capture concentration with $\mu_{\text{ISI}}$ and $\sigma_{\text{ISI}}$. The distribution of the AWGN follows a normal distribution $\mathcal{N}(\mu_N, \sigma^2_N)$. The paper assumes that the AWGN comes from ambient molecules in the environment that is from natural contamination and other molecular emissions. The value used is given in Table I and is taken from [15].

As previously mentioned, the transmission system is a OOK modulation scheme with $\rho$ probability of transmitting a 1. At the optimal sampling time derived previously, the minimum error probability (MEP) criterion of a standard Bayesian detection framework is [17] given by the following

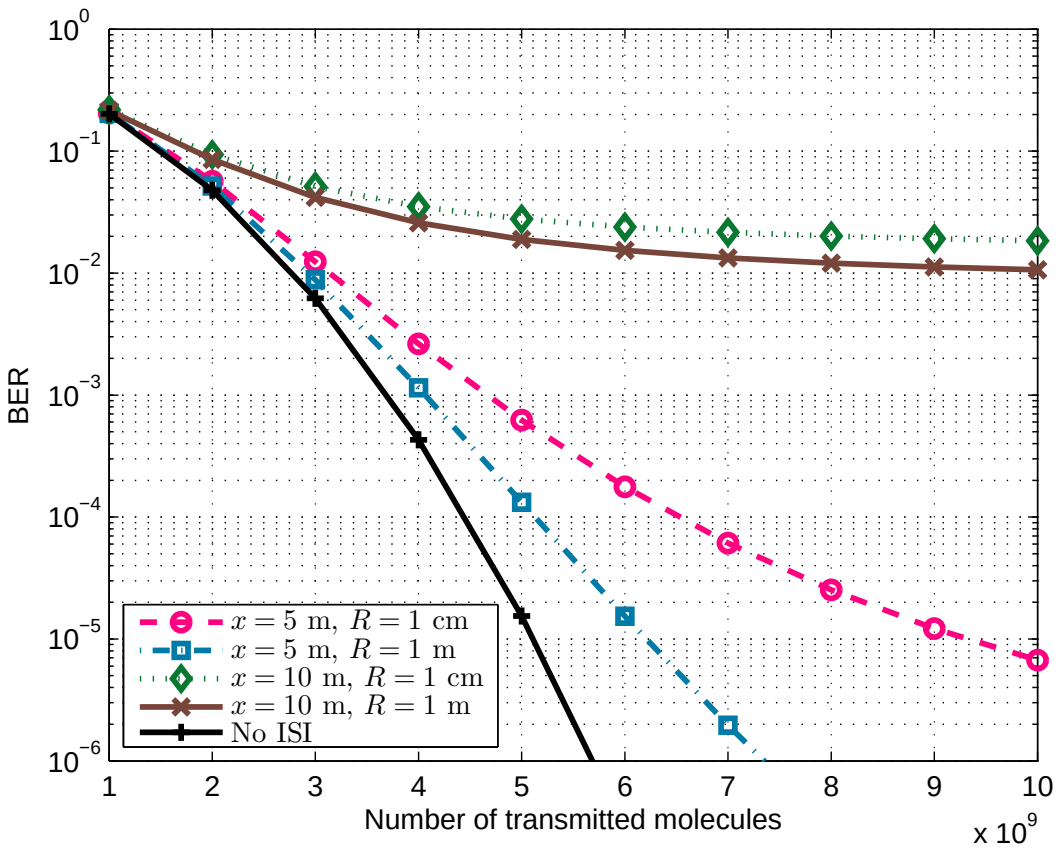

Fig. 4. BER plot for molecular communications, with Gaussian noise and with or without ISI at different transmission distances $x$ and capture zone sizes $R$. The constant parameters is $T_s = 3$ s, $\tau = 0.5$ s, and $v = 5$ cm/s.

with decision threshold $\eta$:

$$\vartheta \gtrless^1_0 \frac{\sigma^2_\vartheta}{\mu_{\vartheta^1} - \mu_{\vartheta^0}} \log\left(\frac{1-\rho}{\rho}\right) + \frac{1}{2}(\mu_{\vartheta^1} + \mu_{\vartheta^0}) \equiv \eta, \tag{11}$$

where:

$$\begin{aligned} \mu_{\vartheta^0} &= \mathbb{E}[\vartheta|\chi = 0] = \mu_{\text{ISI}} + \mu_N, \\ \mu_{\vartheta^1} &= \mathbb{E}[\vartheta|\chi = 1] = \phi_{\max.} + \mu_{\text{ISI}} + \mu_N, \\ \sigma^2_\vartheta &= \text{Var}[\vartheta|\chi = 0] = \text{Var}[\vartheta|\chi = 1] = \sigma^2_{\text{ISI}} + \sigma^2_N. \end{aligned} \tag{12}$$

### D. Bit Error Rate (BER) and Throughput

The average error probability is given by [17]:

$$P_e = \rho Q\left(\frac{\mu_{\vartheta^1} - \eta}{\sigma_\vartheta}\right) + (1-\rho)Q\left(\frac{\eta - \mu_{\vartheta^0}}{\sigma_\vartheta}\right). \tag{13}$$

For line-coding with an equal probability of transmitting a 1 and 0 ($\rho = 0.5$), the BER is reduced to:

$$P_e = Q\left(\frac{\phi_{\max.}}{2\sqrt{\sigma^2_{\text{ISI}} + \sigma^2_N}}\right). \tag{14}$$

Note that the probability of error given a 1 is transmitted is equal to the probability of error given a 0 is transmitted. Therefore, with $\rho = 0.5$, the system is a binary symmetric channel. The achievable throughput of the binary symmetric system is given by [18]:

$$\begin{aligned} C &= 1 - H(P_e) \\ &= 1 + P_e \log_2(P_e) + (1 - P_e)\log_2(1 - P_e). \end{aligned} \tag{15}$$

The paper will now consider a number of macro-scale transmission scenarios and the parameters used to plot the following results can be found in Table I.

*1) Effect of Distance and Capture Zone Size:* In Fig. 4, we demonstrate the BER on a macro-scale ($x$ = 5–10 m), as a function of the number of molecules transmitted M, and various capture radius values $R$. It is clear that the BER is small when only Gaussian noise is considered. When ISI is introduced, the BER deteriorates with increased transmission distance and decreasing capture zone sizes. Each BER with ISI will saturate at a high number of molecules transmitted.

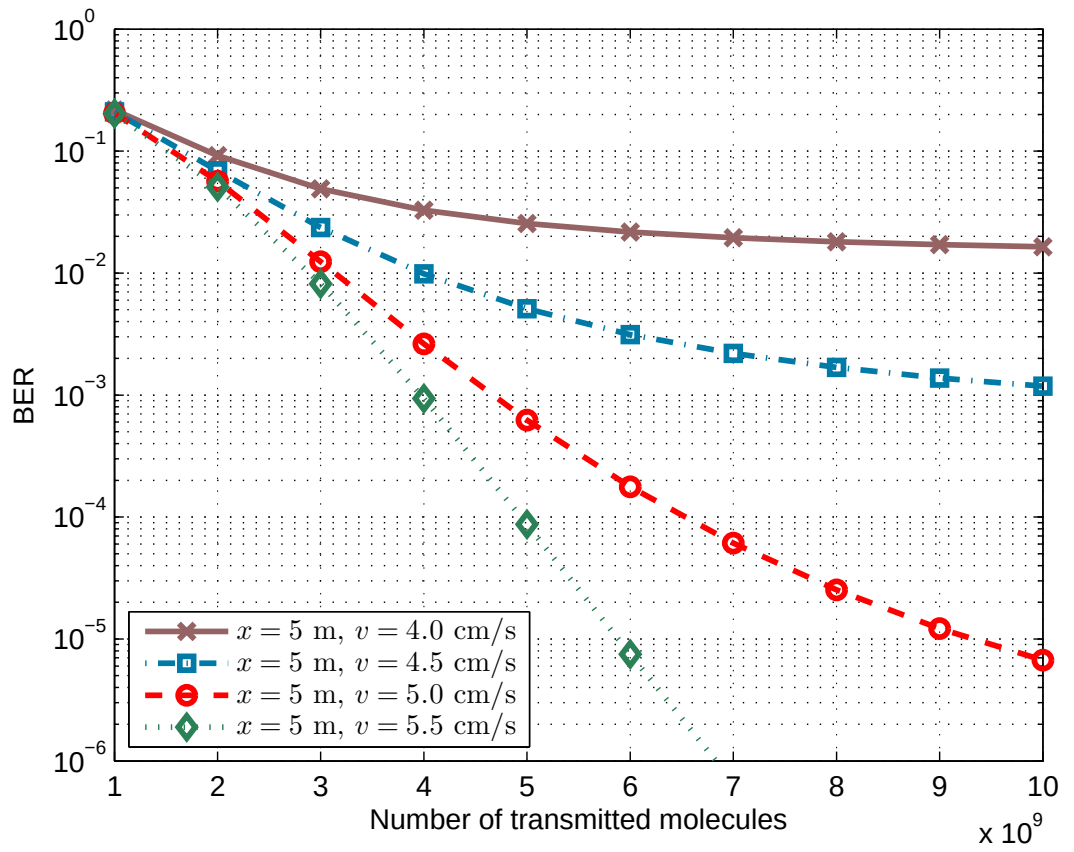

Fig. 5. BER plot for molecular communications, comparing BER with different positive drift velocities $v$. The constant parameters is $T_s = 3$ s, $\tau = 0.5$ s, and $R = 1$ cm.

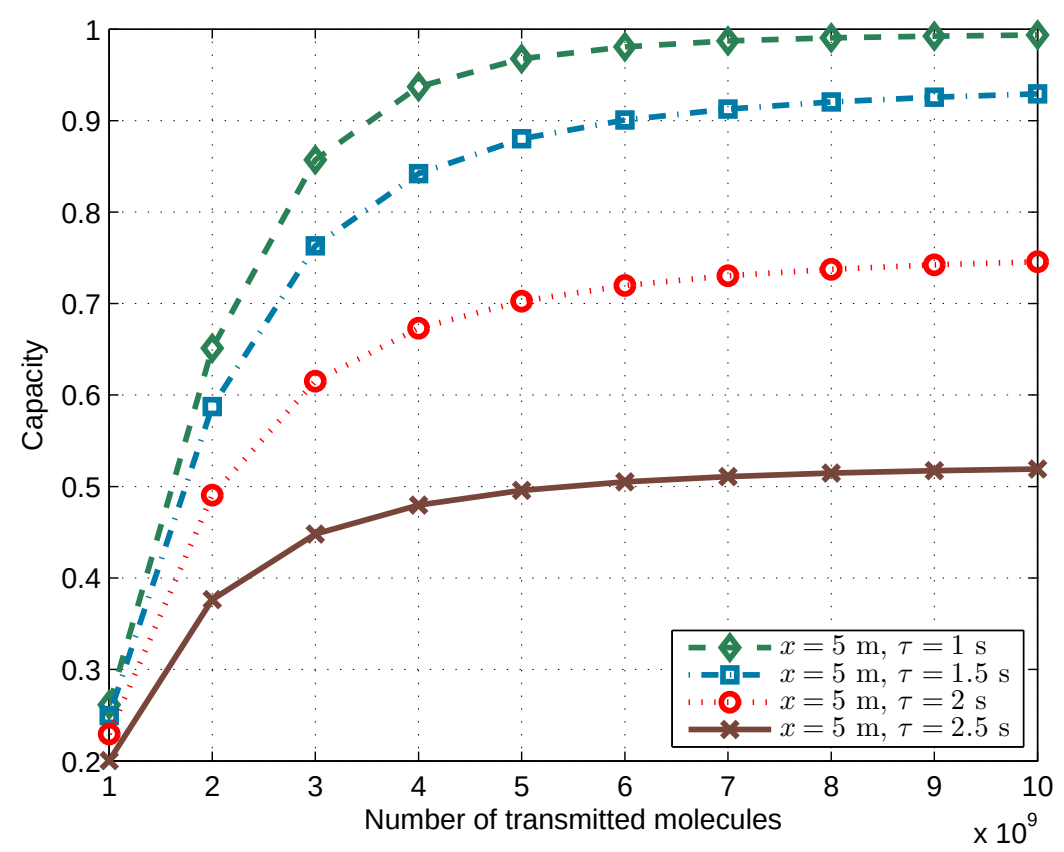

Fig. 7. Throughput plot for molecular communications with Gaussian and ISI noise with different sensor cleanse durations $\tau$. The constant parameters are: $T_s = 3$ s, $v = 5$ cm/s, and $R = 1$ cm.

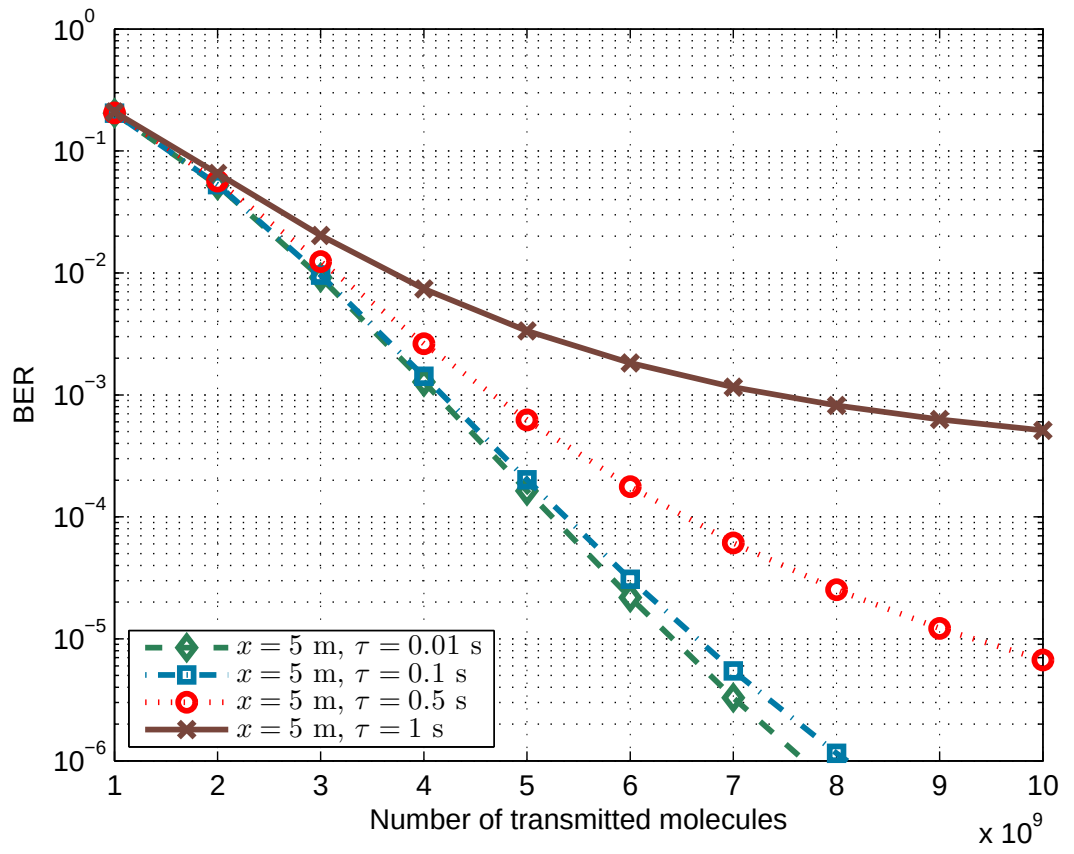

Fig. 6. BER plot for molecular communications with Gaussian and ISI noise with different sensor cleanse durations $\tau$. The constant parameters are: $T_s = 3$ s, $v = 5$ cm/s, and $R = 1$ cm.

TABLE I
SYSTEM PARAMETERS

| *Parameter* | *Symbol and Value* |
|---|---|
| Molecule concentration | $\phi$ |
| Emitted No. of molecules | M, $10^9$–$10^{10}$ mol. |
| Captured No. of molecules | $\theta_c$ |
| Diffusivity | $D$, 0.3 cm$^2$/s |
| Transmission range | $x$, 5–10 m |
| Capture range | $R$, 0.01–1 m |
| Drift velocity | $v$, 4–6 cm/s |
| Optimal sampling time | $t_{\text{max}}$ |
| Maximum molecules detected | $\phi_{\text{max}}$ |
| Captured concentration | $\vartheta$ |
| ISI | $I$ |
| Additive Gaussian noise | $N$ |
| Additive Gaussian noise mean | $\mu_N = \frac{5\times10^8}{R}$ mol./m |
| Additive Gaussian noise variance | $\sigma_N^2 = 0.09\mu_N^2$ |
| Bit error rate | $P_e$ |
| Throughput | $C$ |
| Optimal sampling point | $t_{\text{max}}$ |
| Bit duration | $T_s$ |
| Sensor cleanse time | $\tau$ |
| Round trip time | $t_{\text{RTT}}$ |

*2) Effect of Drift Velocity and Sensor Cleanse Time:* In Fig. 5, we consider low drift velocities in the order of a few cm/s. The BER results include both Gaussian noise and ISI. The plot shows that a small drift current can significantly improve the performance, such that increasing the drift from 4.0 to 5.0 cm/s can reduce the BER by orders of magnitude. Therefore, being able to control and maintain a predictable drift velocity is essential to macro-scale communications.

In Fig. 6, we demonstrate the effect of the sensor cleanse time $\tau$ on the BER in comparison with Gaussian noise. The results show how the BER is very sensitive to the cleanse time, whereby increasing it from 0.1 s to 1 s can increase the BER by several orders of magnitude. Therefore, a rationale conclusion is that the receiver sensor must have a short cleanse time and that the transmission rate must be designed so that it takes into the cleanse time into account.

In Fig. 7, we demonstrate the effect of the sensor cleanse time $\tau$ on the throughput. The results show how the throughput converges to 1 bits/s in the best scenario, and is most sensitive to the cleanse time, whereby increasing it from 1 s to 2.5 s can decrease the throughput from 1 bits/s to 0.52 bits/s.

### E. Macro-Scale System Hardware Design

The design lessons to draw from these results is that several parameters are important to macro-scale molecular communications, and we list them in descending order of importance:

1) Sensor Cleanse Time ($\tau$): a low value can maximise throughput and yield a low BER, with a recommended value of 1 s or below;
2) Drift Velocity ($v$): a positive drift velocity of a few cm/s can significantly reduce the BER and the RTT;
3) Capture Zone Size ($R$): a larger capture zone can reduce the BER, with a recommended value of 10 cm or greater;
4) Turbulent Diffusivity ($D$): greater turbulence will increase diffusivity and shorten the RRT of transmission ($t_{\text{RTT}}$), but also increase the impulse response width, such that the bit duration of transmission needs to be reduced ($T_s$).

The most sensitive parameters are the sensor cleanse time and the drift velocity. Being able to design a receiver with a low cleanse time and designing a transmitter with a controllable drift velocity is critical to achieving a low BER and a high throughput.

## IV. Conclusions

In this paper, we are motivated to use molecular communications to transport data over several metres of distance in order to tackle the challenge of communicating in environments that are hostile to electromagnetic waves. We first derived a novel capture probability expression of a finite sized receiver, with drift. The paper then introduced the concept of time-aggregated molecular noise at the receiver as a function of the speed at which the sensor can self-cleanse. The resulting ISI is expressed as a function of the cleanse time sensor and its effect on the bit error rate and throughput is derived using a Bayesian detector. The results show that the BER and throughput is very sensitive to the sensor cleanse time and drift velocity. The paper goes on to make design recommendations for a macro-scale communication link based on these findings and apply them to a real molecular communication test-bed.

## Acknowledgement

The work in this paper is partly supported by the Australian Research Council (ARC), and the EPSRC Urban Science Doctoral Training Centre at the University of Warwick.